\documentclass[a4paper,12pt]{article}

\usepackage{amssymb,eucal,epsfig,graphicx}

\begin{document}

\begin{center}
{\Huge\bf Gluons in the QCD bound state problem -- a way to exact solution.\\}
\end{center}

\begin{center}
K.A.Ter-Martirosyan\\
ITEP, Moscow, May 22, 2000\\
\end{center}\vspace*{0.9mm}
\begin{abstract}
\noindent The colored objects -- quarks and gluons -- being confined in a small volume $V\sim R_0^3,$ $R_0\sim 0.5$fm inside the QCD bound state get there not small masses $m_{q\bar q}\sim 1$GeV, $m_g\sim 0.5$GeV. This drastically simplifies the QCD dynamics, as now the probabilities e.g. of production of one extra massive valent gluon or extra $q\bar q$ pair, turned to be small – due to a large gap between corresponding energy levels. The ordinary quantum mechanical perturbation theory calculations made in the paper shows that corresponding dimensionless transition amplitudes $\Lambda_{MH}=\frac{V_{MH}}{|E^{(0)}_M-E^{(0)}_H|}$ (from meson $q\bar q$ to the hybrid $q\bar qg$) have a value of 20--25\% and the mixing of hybrid with glueball $(gg)$ is even smaller (of the order of 10--18\%). Taking into account all such mixings with all highly lying Fock states is equivalent to the exact solution of QCD bound state problem. This can be done really summing the small perturbation theory results for these states contributions.\\
\end{abstract}
{\bf 1. INTRODUCTION/MOTIVATION }\\

$\bullet$ Massless quarks and gluons acquire inside of hadrons not small constituent masses

$$
 m_q\simeq\frac{1}{R_0}\simeq0.35\mbox{GeV},\quad m_g\simeq0.45 \mbox{GeV}
$$

\noindent due to gluon string confinement in a small volume $V_0\sim R_0^3,\; R_0\simeq 0.5-0.6$fm.
Also the gluon strings themselves are massive  having about the same mass $M_{string}\sim 0.4-0.5$GeV, so $m_g \sim M_{string}$ at string tension of about 0.18-0.15GeV$^2$.

$\bullet$ Therefore each extra gluon, or a $q\bar q$ pair of quarks lead to enhancement of hadron mass to about $ 0.8-1.0$GeV.
This was confirmed by lattice calculations.

$\bullet$ These estimations can be illustrated comparing  e.g. the $\rho$ --meson mass $m_\rho\simeq0.8 $GeV$\; (\rho=(\bar qq)^{s=1})$
with the mass of vector hydrid $ h = ( \bar qq, g ) ^ {s=1} )$
for which $ m_h \simeq 1.8$ GeV. Another example represents glueballs, $G=(gg)$, consisting from two valent (constituent) 
gluons bound by a pair of gluon strings and observed apparently with the mass $m_G\simeq 1.5-1.8$GeV.
These recent data agree well with the estimations given above and with lattice calculation data. 


\renewcommand{\footnoterule}{\vspace*{-3pt}\hrule width .8\columnwidth \vspace*{2.6pt}}

$\bullet$ This situation can naturally be understood dividing the gluon field in two parts (as t'Hooft and Simonov [1] have suggested):               
$ \hat A_\mu (x)=\hat B_\mu (x)+\hat a_\mu(x)$ ,
where $\hat B_\mu(x)=t_\alpha B_\mu^\alpha (x)$ leads to gluon string formation and the part
\begin {equation}
\hat a_\mu(x)=t_\alpha a_\mu^\alpha (x)={\sum_{\vec k, \alpha}}'\frac{\hat t_\alpha}{\sqrt{2\omega_k}}
[c_\mu^\alpha(\vec k)e^{+i\vec k\vec x}+{c_\mu^\alpha}^+(\vec k)e^{-i\vec k\vec x}]
\end {equation}
is responsible for the valence (constituent) gluon production and absorbtion.

The QCD interaction energy has the form
$$
\hat V_{QCD}=\hat v'_{QCD}(\hat a_\mu)+\hat V'_{QCD}(\hat B_\mu)+\hat V'_{\hat a_{\mu\nu}, \hat B_\mu}.
$$
where term
\begin{equation}
\hat v'_{QCD}(\hat a_\mu)=g_s\int [(\bar q (x)\gamma_\mu t_\alpha q(x))a_\mu^\alpha (x)-
\frac{1}{2}f_{\alpha\beta\gamma}a^\alpha_\lambda a^\beta_\sigma(\partial_\lambda a_\sigma^\gamma-\partial _\sigma a_\lambda^\gamma )-O(\hat a_\mu^4)]d^3x
\end {equation}
changes the numbers $ n,\; N$ of valence gluons and of $q\bar q$ pairs respectively;
the term $\hat V'_{QCD}(\hat B\mu)$ leads to QCD strings production and to colour confinement, while $V'_{\hat a_\mu \hat B_\mu}=\sum\limits_{k=1}^3c_k(\hat a\mu)_k(\hat B\mu)^{4-k}$ is not important below.\\

{\bf 2. THE ZERO ORDER HAMILTONIAN AND BOUND STATES IN THE SEPARATED QCD SECTORS.}\\

$\bullet$ Considering $\hat v'_{QCD}$ as perturbation let us introduce the zero order QCD Hamiltonian $\hat H^o_{QCD}$:
\begin {equation}
\hat H_{QCD}=\hat H^0_{QCD}+v'_{QCD}(a_\mu)
\end {equation}
with the complete and orthogonal system of its eigenfunctions $\Phi ^{(n)}_N$\\ 

\vspace*{-0.5cm}
\begin {equation}
\hat H^0_{QCD}\Phi ^{(h)}_N=E^{(n)}_N\Phi ^{(n)}_N
\end {equation} of separate Fock sectors with a fixed numbers $n,\; N$ of gluons and of $q\bar q$ pairs (at $v'_{QCD}=0$ their numbers $n,\; N$ remain fixed!) Both the QCD Hamiltonian $H_{QCD}$ (3) and its zero order 
form\footnote[1]{which includes only the string interaction part $V'_{QCD}(B_\mu)$ of the total QCD interaction.}
 $H^0_{QCD}$ were introduced and intensively used in papers by Simonov and collaborators [1,2,3].Their results are largerly used below. Let us stress that having the values of all mixing rates $|\lambda _j^{(k)}|^2$ i.e. of  all $C_j^{(k)}= C_j \lambda _j^{(k)}$,one directly recovers using Eq.(5) below the exact QCD solution $\Psi_j $ of bound state problem. It is explained farther that $\lambda _j^{(k)}$ can be found  in the form of usual quantum mechanical theory rapidly convergent series considering QCD interaction $v'_{QCD}$ of constituent quarks and gluons in Eg. (3) as small perturbation.


$\bullet$ The exact solution $\Psi_j$ of QCD bound states problem can be written in general as superposition of the zero order eigenfunctions $\Phi^{(n)}_N$ of Hamiltonian $H^o_{QCD}$:
\begin{equation}
\Psi_j=\sum_{N,n=0}^{\infty}C_{N,n}^{(j)}\Phi _N^{(n)},\qquad\mbox{ with }\quad\sum_{n,N}|C_{n,N}^{j}|^2=(\Psi_j^*\Psi_j)=1
\end{equation}
as $(\Phi _N^{(n)}\Phi_N'^{(n)*})=\delta_{n,n'}\delta_{N,N'}$. Here $j$-- can be the meson $j=M\; (\bar q q)$, the glueball $j=G=(gg)$, or different hybrids: $j=H=(q\bar q g),\mbox{ or } j=H'=(q\bar q2g)$ etc.

$\bullet$ It is usefull to put the superpositions (5) for $j=M,G,H$ in the form of Fock columns. 
\vphantom{$
\begin{array}{c}
A\\A\\A\\A\\A\\A
\end{array}$}
Denoting $N,n=k=(M,G,H),\; C_j^{(k)}=C_j\lambda^{(k)}_j$ one has respectively:
\begin {equation}
\Psi_M=C_M
\left(
\begin{array}{c}
\Phi _M\\
\lambda_M^{(H)}\Phi_H\\
\lambda_M^{(G)}\Phi_G\\
...\\
...\\
\end{array}
\right), \quad
\Psi_G=C_G
\left(
\begin{array}{c}
\Phi _G\\
\lambda_G^{(H)}\Phi _H\\
\lambda_G^{(M)'}\Phi _M\\
...\\
...\\
\end{array}
\right), \quad
\Psi _H=C_H
\left(
\begin{array}{c}
\Phi _H\\
\lambda_H^{(M)}\Phi _M+\lambda_H^G \Phi _G \\
...\\
...\\
...\\
\end{array}
\right)
\end {equation}
where $C_M=(1-|\lambda^H_M|^2-|\lambda^{(G)}_M|^2)^{1/2}\simeq 1\simeq C_G\simeq C_\mu$ and all
$\lambda ^{(k)'}_j\ll \lambda ^{(k)}_j \ll 1$ as the mixings with higher Fock states $\lambda ^{(k)}_j=\frac{V_{jk}}%
{E^{(0)}_j-E^{(0)}_k}\ll 1$ are proved to be small $(\lambda ^{(k)}_j \sim 10-20 \%)$ -- see below.\\[0.6cm]



{\bf 3. THE QUANTUM MECHAMICAL (QM) PERTURBATION APPROACH (PA).}\\

$\bullet$ Now we are ready to apply the usual QM perturbation theory (PT) which gives the well known results:
\begin {equation}
\left \{ \begin{array}{l}
E_n=E_n^{0}+\sum\limits_{k \ne n}\frac{|V_{nk}|^2}{E_n^{0}-E^{0}_k}+ 
\sum\limits_{k,l\ne n}\frac{V_{nk}V_{kl}V_{ln}}{(E^0_{n}-E^0_{k})(E^0_{n} -E^0_{l})}+...\\
\hspace*{-5mm}\\
\Psi_n=\Phi _n[1-\frac{1}{2}\sum\limits_{k\ne n}\frac{V_{nk}V_{kn}}{(E_n^{(0)}-E_k^o)}^2] +\sum\limits_{k \ne n}\Phi _k\frac{V_{nk}}{E^{0}_n-E^{0}_k }                                                                                                                                                                                                                                                                                                                                                                                                                                 + (1-\frac{V_{nk}}{E_n^0-E_k^0})             
 \sum\limits_{k,l\ne n}\Phi _l\frac{V_{lk}V_{kn}}{(E^{0}_{n}-E^{0}_{k})(E^{0}_{n} -E^{0}_{l})}+...\\
\end{array}
\right. 
\end{equation}
\vspace*{-0.5cm}where 
\begin {equation}
V_{kn}=\int\Phi _n^*\hat v'_{QCD}\Phi _kd\tau
\end {equation}
and operator $\hat v'_{QCqD}$ transists $\Phi _M\rightleftarrows \Phi _H\rightleftarrows \Phi _G$, so that $V'_{kn}\ne 0$ only for $k\ne n$

$\bullet$ Let us put the zero order wave functions $\Phi _n^j$ for $j^{PC}=0^{++}$ and $1^{- -}$ mesons in the form:\\
\begin {equation}
 \left \{ \begin{array}{l}
(\Phi ^j_M)^{\lambda \sigma}_{\alpha \beta }=(R_M^j)_{\alpha \beta }^{\lambda \sigma,}\varphi _M^j(\vec r_{12}),\; \\
\hspace*{-3mm}\\
(\Phi ^j_H)^{\lambda \sigma}_{\alpha \beta }=(R_H^j)_{\alpha \beta ,\mu }^{\lambda \sigma ,\alpha  }\varphi _H^j(\vec r_{12},\vec r_{13}),\;\\
\hspace*{-3mm}\\
(\Phi^j_G)_{\mu\nu }^{ab}=(R_G^j)^{ab}_{\mu,\nu }\varphi _G^j(\vec r_{gg})\\
\end{array}
\right.
\end {equation}
separating vertices $R^j_A$ of meson decays into quarks and gluons (shown in Fig.1,2) from the radial parts $\varphi_M (\vec r_{12}),\,  \varphi _H(\vec r_{12},\vec r_{13}),\,\varphi _G(\vec r_{gg})$ of the corresponding wave functions; here $a,b=1,2,...,8$, $\lambda =1,2,3$ are gluon and quarks colour states indices wile $\mu,\nu =1,2,3,4$ are gluon Lorents indices and $\lambda ,\beta =1,2,3,4$ are Dirac quarks indices.

The explicit forms of vertices for $0^+$ and $1^-$ cases are:
\begin {equation}
\begin{array}{ll}
\left\{
\begin{array}{l}
(\hat R^{0^+}_M)^{\lambda \sigma }_{\alpha \beta }=\frac{1}{\sqrt{12}} \delta^{\lambda \sigma }\delta_{\alpha \beta }\\
\hspace*{0.9mm}\\
(\hat R^{0^+}_H)^{\lambda \sigma, a }_{\alpha \beta ,\mu }=\frac{1}{4\sqrt 2}t_a^{\lambda \sigma }e_\mu^a(\gamma_\mu)_{\alpha \beta }\\
\hspace*{1mm}\\
(\hat R^{0^+}_G)^{ab}_{\mu \nu }=\frac{1}{4\sqrt 2}\delta^{ab}g_{\mu\nu}\\
\end{array}
\right.&
\left\{
\begin{array}{l}
(\hat R^{1^-}_M)^{\lambda \sigma }_{\alpha \beta }=\frac{1}{4}\delta^{\lambda \sigma }(\gamma_\mu)_{\alpha\beta}\\
\hspace*{1mm}\\
(\hat R^{1^-}_H)^{\lambda \sigma}_{\alpha \beta ,\nu }=\frac{1}{4}t_a^{\lambda \sigma }e_\mu^a(\gamma_\nu\gamma_\mu)_{\alpha \beta }\\
\hspace*{1mm}\\
(\hat R^{1^-}_G)^{ab}_{\mu \nu }=\frac{1}{4\sqrt 2}\epsilon_{\mu\nu\lambda \sigma }e_\nu^ae_\lambda^bi\vec\bigtriangledown_\sigma\\ 
\end{array}
\right.
\end{array}
\end {equation}
they are normalized to $1:\; (R_A^jR_A^{j*})=1 $ for all $A=M,H,G$, where  $a=1,2,...,8$ and $\hat t_a=\hat\lambda_{a/2}$,  are Gell--Mann $3\times 3$ matrices.

$\bullet$ With these zero order wave functions the QCD perturbation operator $v'_{QCD}$ (see the first term in Eq.(2)) can be used in calculation of $V_{kn}$ directly in $\vec r-$space in the following "stripped" form:
$$
\hat v'_{QCD}=(g_s/\sqrt{2\omega_c})(\gamma_\mu\hat t_ce_\mu^c)[\delta^{(3)}(\vec r_{13})+\delta^{(3)}(\vec r_{23})]
\eqno (10a)
$$
as quarks $q,\bar q$ and gluons $g$ individual wave function can be absorbed (in Eq.(8) for $V_{kn}$) into $\Phi_k$ and $\Phi_n$ while $\delta^{(3)}$ terms prove to appear in $(v'_{QCD})_{stript}$ as e.g. transition $M\to H$ is reached by production of a gluon by quark or anti quark at some point $\vec r=\vec r_3$ and also by absorbtion and production of quark essentially at the same point $\vec r = \vec r_1$, i.e. at $\vec r_{13}=0$ (or at the point $\vec r=\vec r_2$, i.e. at $\vec r_{23}=0$ for anti quark). For $H\to G$ transition the situation is simpler and here the action of $v'_{QCD}$ involves the appearing of $\delta ^{(3)} (\vec r _{12})$: \\
$$
(v'_{QCD})'_{stript} = g_s/ \sqrt {2 \omega _c}(\gamma _\mu \hat t_c e^c_\mu )\delta ^{(3)} (\vec r_{12})
\eqno (10b)\\
$$
due to a locality of the QCD interaction; see the first term in Eq.(2) (acting on the hybrid state it absorbs $\bar q,q$ in some point $\vec x= \vec x_1$ and produce gluon $g$ just in the same point).

This "stripped" version of PT follows simply from the momentum representation (like one used by Orsay group paper [4]) of all constituent gluons and quarks operators, e.g. for quarks
$$
q_\alpha (x)=\sum_{\vec p,\sigma ,\lambda }\frac{1}{\sqrt {2\epsilon (\vec\rho )}}[u_\alpha (\vec p,\sigma )v_\lambda e^{i\vec p\vec r}+ u_\alpha^c(\vec p,\sigma)v_\lambda^+e^{i\vec p\vec r}]
$$ 
and similarly for gluons by the above determined ope\-ra\-tor $\hat a_\mu(x)$.

Authors of two recent publication [6,7] (from North Carolina Univ.: E.Gubankova, \mbox{C.--R. Ji} and S.R.Cotanch) have calculated the glueball spectrum and obtained the value of gluon mass $m_g\simeq 0.5$GeV inside of confinement region -- close to the Simonov results [1] used here. They proposed to consider the colour Coulomb interaction of quark and gluons as the sours of  mixing between different states in each QCD isolated sectors (with a given numbers $n_g$ of gluons and  $N=N_{q\bar q}$ of valent $q\bar q$ pairs). Let us remind that t'Hooft identity
\footnote[1]{Physically it means that gluon strings stretched between valent quarks or gluons can have the transversal oscillations. After quantisation the quants of these oscillations will represent the valence gluon states.

The potential $v'_{QCD}(a_{\hat\mu})$ determines their interaction whith quarks and leads to production or absorption of  these valence states of gluons and of $q\bar q$ pairs.} 
($A_{\mu}(x)=B_{\mu}(x)+a_{\mu}(x)$, see above) allowed us to put the QCD interaction in the form $\hat V_{QCD}=\hat V'_{QCD}(\hat B_\mu)+v'_{QCD}(a_\mu)+\cdot\cdot\cdot$, with $\hat V'_{QCD}(\hat B_\mu)$ responsible for the binding  of all separate Fock state $\Phi^{(n)}_N$ (with a given $n_g,\; N$ see Eq.4).

However, a short disqussion made above has clearly shown that, while the Coulomb coulor interactions of constituents inside of each sectors can be important as authors of Refs.[6.7] notes (less important, however than the string type one), the real cause for mixings of bound states is the QCD interaction (2) $v'_{QCD}(\hat a_\mu)$ of valence gluons and quarks. It changes the numbers of them converting e.g. the mesons ($n_g=0,$ $n_{q\bar q}=1$) into hybrid ($n_g=n_{q\bar q}=1$) and in glueballs ($n_g=2,$ $N_{q\bar q}=0$) and really mixing them. Just the rates of these mixings are calculated below.

$\bullet$ The correction (7) to $E_n$ and to $\Psi_n$ of PT determined above proves to be not too small-of the 
order of 10--20\%, i.e.$\lambda_A^{(B)}=\frac{V_{AB}}{E^0_A-E^0_B}\lesssim 0.10-0.20,\mbox{ for } A,B=M,H,G$. Therefore one can calculate them using approximate zero order wave functions. The explicit form of meson and glueball wave functions can be
determined from Schroedinger equation with the string type linear rising potential $V_{string}\simeq\sigma|\vec r|,\,\sigma\simeq0.18$GeV$^2$, of $q\bar q$ interaction or respectively, by the potential $V_{gg}\simeq (9/4)\sigma |\vec r|$ for gluon-gluon interaction case. They can be approximated (with high accuracy) by a simple Gaussians:
\begin {equation}
\left \{ \begin{array}{l}
\varphi _M(r)=(\frac{3}{2\pi r_0^2})^{3/4}e^{-\frac{3}{4}(r/r_0)^2},\\
\hspace*{-5mm}\\ 
\varphi _G(r)=(\frac{3}{2\pi r_{g0}^2})^{3/4}e^{-\frac{3}{4}(r/r_{g0})^2}
\end{array}
\right.
\end {equation}
whith the proper chosen values of $r_0^2$ and $r_{g0}^2$ which are 
$$
\begin{array}{c}
r_0^2=\langle \vec r^2\rangle_M=0.725fm^2,\, r_o \simeq 4.32(GeV)^{-1}\\
\hspace*{-5mm}\\ 
r_{g}^2=\langle \vec r^2\rangle_{g}=0.32fm^2,\, r_{g} \simeq 2.88(GeV) ^{-1}\\
\end{array}
$$
respectively  obtained at $\sigma=0.18$GeV$^2, $and at $  \sigma _{gg}=\frac{9}{4}\sigma $.

$\bullet$ The hyperspherical formalism [4,1] 
allows one to obtain with the same accuracy the three--particle ($q_1,\,q_2$ and $g$) hybrid's wave function
$\varphi _H(\vec r_{12}\vec r_3)=\rho^{-5/2}\chi_H(\rho)u_k(\Omega_\sigma)$ ,
where $u_k(\Omega_\sigma)=const$ for the hybrid ground state, and\\ $\rho^2=\vec \xi^2+\vec \eta^2=\sum\limits_{i,k}\frac{\mu _i \mu _k}{2m\mu}r_{ik}^2$ with $\mu =\mu _1+\mu _2+\mu _3=2\mu _1+\mu _3$ at $\mu_1= \mu_2= \mu_q,\mu_3=\mu_g$.\\ At  $r_{13}=0$ one finds $\vec r_{13}=\vec r_{12}$ and  $\rho^2=\frac{\mu_1(\mu_2+\mu_3)}{2m\mu}\vec r^2_{23}$ where $ \mu+\mu_3=2(\mu_1+\mu_3)$ and the scale $ m $ can be choosen arbitrary. Putting $ \mu \cdot m_0=\mu_1(\mu _1+\mu_3) $ or $m_o =\mu _1\frac{\mu _3+\mu_1}{\mu _3+2\mu _1}$ one obtains at $r_{13}=0$:
 $[\rho ^2]_{r_{13}=0}=r_{13}^2=r_{12}^2=r^2$.

\hspace*{0cm}
\begin{picture}(15,15)
\put(101,11){\vector(-2,-1){20}}
\put(80.5,0.5){\circle*{1}}
\put(101,11){\circle*{1}}
\put(78,-7){$1$}
\put(100,13){$2$}
\put(91,6){\vector(3,-1){25}}
\put(116,-2.5){\circle*{1}}
\put(87,8){$\vec\eta$}
\put(100,-6){$\vec\xi$}
\put(118,-5){$3$}
\put(89,-15){$\vec\rho^2=\vec\xi^2+\vec\eta^2$}
\end{picture}\\

\vspace*{15mm}
Note that at\quad $r_{12}=0$\quad $\rho^2=(\frac{2\mu_3}{\mu_1+\mu_3})^{1/2}r_{13}^2$\quad (here $r_{13}\equiv r_{23}$),\qquad i.e. $\rho= a_0r_g$ with\\ $a_0=[2/(1+\mu_1/\mu_3)^{\frac{1}{2}}]^{1/2}\simeq 1.083$ at \mbox{$\mu_1=\mu_q=0.312$GeV,} $\mu_3=0.44$GeV,  $m_o= \mu_1\frac{1+\mu _1/\mu _3}{1+2\mu _1/\mu _3}$\\ (in the glueball $\mu _3=(\mu _g)_G=0.53$GeV, while inside of hybrids [1] $\mu_3=(\mu_g)_H\simeq0.44$GeV).

In the hyperspherical method the total potential [4,1]
\begin {equation}
W(\rho)=\varkappa_0\sigma \rho +\frac{L(L+1)}{2m\rho^2}
\end {equation}
(where $L=\frac{3}{2}+K,\; K=0,1,...;\; L=\frac{3}{2}\mbox{ at }K=0$ and $\varkappa_0=\eta_0 \frac{\mu_1+\mu_3}{\sqrt{\mu\mu_3}}=\eta _0\frac{1+\mu_1/ \mu_3}{\sqrt{1+2\mu_1/ \mu_3}},\;$ $\eta _0=\sqrt[3]{5/2}=1.357$) has a minimum at $\rho=\rho_0\sim1/(m\sigma)^{1/3}$ where $W'(\rho)=0$, i.e. at: $\varkappa  _0\sigma =\frac{L(L+1)}{m\rho _0^3}$, (see Fig.5.), resulting in:


\begin{center}
\begin{tabular}{p{0.6\columnwidth}p{0\columnwidth}|p{0.3\columnwidth}|}
\cline{3-3}
\begin{equation}
\rho_0=\frac{1}{\sqrt m}(\frac{\mu_1\mu_3}{\mu_1+\mu_3})^{1/6}(\frac{L(L+1)}{\varkappa_0\sigma})^{1/3}
\end{equation}
& &
 \\[10mm]
Therefore for the lower hybrid state one obtains practically the oscillation potential (see Fig.5):
& & 
\\[6mm]
\begin{equation}
\begin{array}{c}
\quad  W(\rho)\simeq W_0(\rho)=\frac{m\omega_0^2}{2}(\rho-\rho_0)^2\\[5mm]
\mbox{with\quad }\omega_0^2=W''(\rho_0)=\frac{3L(L+1)}{(m\rho _0^2)^2},\quad\omega_0= \frac{\sqrt{3L(L+1)}}{m_0\rho ^2}\\[5mm] 
\mu_1=\mu_2=\mu_q,\; \mu_3=m_g,\; \mu=2m_q+m_g
\end{array}
\end{equation}
& & \\
\noindent This gives the normalized hybrid wave function in the following simple Gaussian form similar to Eq.(10): 
& & \\
\cline{3-3}
\end{tabular}
\end{center} 

\vspace*{-86.5mm}
\hspace*{107mm}\includegraphics[width=53.5mm, height=81.4mm]{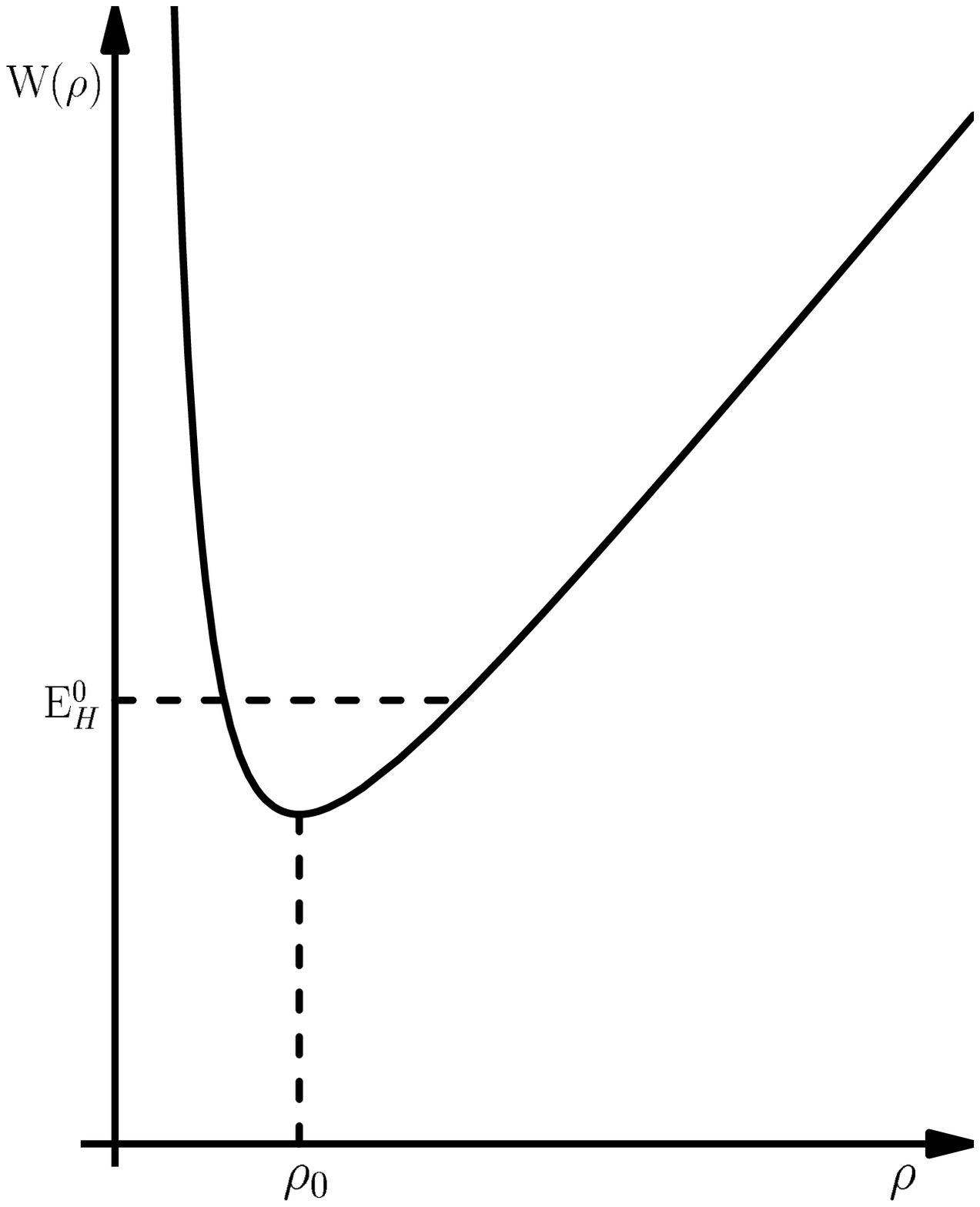}

\vspace*{2mm}\hspace*{128.5mm}{\bfseries Fig. 5}

\vspace*{-6.5mm}
\begin {equation}
\varphi _H(\vec r_{13}\vec r_{23})=\frac{|det|^{-1/2}}{\sqrt{\Omega_6}}(\frac{m\omega_0}{\pi})^{1/4}\rho^{-5/2}  e^{-\frac{m\omega_0}{2}(\rho-\rho_0)^2}
\end {equation}
where $\frac{1}{\rho^{5/2}}=(\frac{1}{\rho_0^{5/2}})[1+\frac{5}{2}(1-\frac{\rho}{\rho_0})+...]$ will be used for the effectively small $\Delta\rho =\rho-\rho_0\ll \rho _0$, \\
\raisebox{-20pt}{ while}

\vspace*{-20pt}
$$
\Omega_6=\int d\Omega_6=\pi^3,\qquad |det|^{1/2}=
|det\left(
\begin{array}{ccc}
\vec r_1&\vec r_3 & \vec r_3\\
\vec\xi&\vec\eta &\vec R 
\end{array}\right)|^{1/2}=
\left(
\frac{\mu_1\mu_2\mu_3}{m^2\mu}
\right)^{3/4}
$$
and the normalization means
$$
\begin{array}{c}
 \frac{1}{V}\int d^3r_1d^3r_2d^3r_3|\Phi_H|^2=\int d^3\xi d^3\eta|det||\phi_H|^2=\quad
\hspace*{-5mm}
=\int |det|\rho^5d\rho d\Omega_6 |\phi_H|^2=1\\
\end{array}
$$

\hspace{6mm}\\
{\bf 4. MATRIX ELEMENTS OF MH,HG,MG; MIXINGS AND CORRECTIONS TO PARTICLE BOUND STATES ENERGIES (MASSES)}\\

$\bullet$ Matrix elements $V_{MH},\; V_{HG},\; V_{MG}$ are generically small -- of order $\sim(1/5)-(1/10)$ GeV; let us show this for the simplest $0^{++}$ mixings case:\\
\hspace*{2cm} a) The meson-hybrid mixing amplitude is represented in the first PT approximation by two graphs 3a,3b,yielding:
$$
V_{MH}=
 2\frac{g_s}{\sqrt{2\omega_0}}\int \varphi ^*_M(r)(\hat R^{0^+}_M(\hat \gamma_\nu e_\mu^c\hat t_c)\hat R^{0^+}_H)\varphi _H(\vec r)d^3r
=\sqrt{\frac{8}{3}}\frac{g_s}{\sqrt{\omega_0}}\int\varphi ^*_M(r)\varphi _H(r,r_{13}=0)d^3r\, ,
$$
$$
\begin {array}{c}
 (\hat R^{0^+}_M(\hat \gamma_\mu e_\mu^a\hat t_a\hat)R^{0^+}_H)=\frac {2}{\sqrt{3}}.\\
\end {array}
$$
\raisebox {1cm}[0pt]{as}
\begin{equation}
 V_{MH}=A_{MH}J(a,b)
\end{equation}
\raisebox{1cm}[1pt]{Thus}
\raisebox{0cm}[0pt]{\hspace*{-1cm}where}
\begin{equation}
A_{MH}=\frac{2^{13/4}}{\pi^{3/2}}\frac{ g_s}{ r_0}    \frac{(1+\mu_1/\mu _3 )^{7/4}}{1+2\mu _1/\mu _3}(\frac{\mu _1}{\omega _0})^{1/4}\, , 
\end{equation}
\raisebox{0cm}[0pt]{while}
\begin{equation}
J(a,b)=1/b^{5/2}\int\limits_0^\infty e^{-t^2-a(t-b)^2}t^2dt[1+\frac{5}{2}(1-t/b)+...], \quad
t=\frac{\sqrt{3}r}{2r_0}
\end{equation}
and the decomposition of $\varphi  _H$ in power of $\frac{\rho -\rho _0}{\rho _0}=\frac{t-b}{\rho_0}\ll 1$ (shown above after Eq.(15)) was used, so that $\varphi_H(r_{13}=r,r_{23}=0)\sim t^{-5/2}e^{-a(t-b)^2}$ was written at $|t-b|\sim 1/a<b$ as: \\
$b^{-5/2}e^{-a(t-b)^2}[1+\frac{5}{2}(1-t/b)+ \frac{35}{8}(1-t/b)^2+...]$ 
with dimensionless $a=\frac{2m_0\omega_0}{3}r_0^2,\; b=\frac{\sqrt{3}\rho_0}{2r_0},$ \\ with $\rho_0,\; \omega_0=(\sqrt{\mu_1^2+\vec k_0^2})_{eff}\;$ determined in Eqs.(13),(14) through quark and gluon masses  $\mu_1,\;\mu_3,$ resulting in $\rho_0\simeq 3.86\mbox {GeV$^{-1}$},\;\omega_0\simeq 1.02\mbox {GeV}$.

The simple numerical calculations yield: $a\simeq2.79,\; b\simeq 0.775,$ and $A_{MH}\simeq 0.694$GeV obtained at $g_s=2.24, \; \alpha_s\simeq0.4,$ while for the rapidly convergent integral (18) (where the integrand in  $J (a,b)$ is shown in Fig.6a) one obtains: $J(a,b)=0.290$ in the first approximation. This results in:
$$
V_{MH}=A_{MH}J(2.79,0.775)\simeq 0.694{\mbox{GeV}}\cdot 0.290=0.201\mbox{GeV},
$$ 
leading to about 22\% for MH dimensionless mixing amplitude: $\lambda_{MH}=\frac{V_{MH}}{|E^0_m-E^0_H|}\simeq 0.22$ for $|E^0_m- E^0_H| \simeq 0.8 GeV$.\\
\hspace*{2cm} b) Quite similarly for hybrid -glueball mixing one has (see two graphs in Fig.4, each giving the half of $V_{HG}$ value ):
$$
\begin{array}{c}
V_{HG}=
\frac{g_s}{\sqrt{2\omega}}\int \varphi_H^*(c_0 r,r_{12}=0)(\hat R^{0^+}_H(\hat \gamma_\nu e_\nu^c\hat t_c)\hat R^{0^+}_G)\varphi_G(r)d^3r=\\ \\
=\frac{g_s/\sqrt{2}}{\sqrt{2\omega_0}}\int\varphi^*_G(r)\varphi_H(c_0 r)d^3r=A_{HG}\tilde J
\end{array}
$$
\begin{equation}
\mbox{with}\hspace*{2cm} A_{HG}=\frac{1}{2}\sqrt{\frac{3}{8}}\frac{r_0}{r_{g}}A_{MH}=\frac{\sqrt{3}\cdot{2}^{3/4}g_s}{\pi^{3/2}r_{g}}\frac{(1+\mu_1/\mu_3)^{7/4}}{1+2\mu_2/\mu_3}(\frac{\mu_1}{\omega_0})^{1/4}\simeq 0.320\mbox{GeV}
\end{equation}
\begin{equation}
\mbox{\raisebox{0.75cm}[0pt]{and}}\hspace*{-0.5cm}
\tilde J= J(\tilde a,\tilde b)=\tilde b^{-5/2}\int\limits_0^\infty e^{-t^2-\tilde a(t-\tilde b)^2} t^2dt \mbox [1+\frac{5}{2}(1-\frac{t}{b})+\frac{35}{8}(1-\frac{t}{b})^2+...],\quad t=\frac{\sqrt 3}{2r_g}a_0r
\end{equation}
as here $(\hat R^{0+}_H(\hat \gamma_\nu e_\mu^b\hat t_b)\hat R^{0+}_G)=1/ \sqrt{2}$ and one had $\rho=c_0r$, $c_0=\sqrt{\frac{2\mu_3}{\mu_1+\mu_3}}\simeq 1.083$ at $r_{12}=0$. \\
Also: as $\frac {c_0 r_g}{r_0}\simeq 0.719$ then:
$$
\left\{
\begin{array}{l}  
\tilde a= \frac{2m_o\omega_o }{3}(c_o r_g)^2 =(\frac {c_o r_g}{r_o})^2a\simeq 0.517a \simeq 1.444\; \\
\hspace*{-3mm}\\
\tilde b=\frac{\sqrt{3}\rho _0}{2c_0 r_g}=b/(\frac{c_0r_g}{r_0})\simeq b/0.719=1.078
\end{array}
\right.
$$
Simple calculations of $\tilde J=|J(\tilde a,\tilde b)|$(see Fig 5.b)  gives: $\tilde J\simeq 0.285$, and thus:
$$
V_{HG}=A_{HG}\tilde J=0.320GeV\cdot 0.285=0.091 GeV
$$ 
with the first correction in Eq.(20) at $A_{HG}=0.320$. 
That results to about 18\% of H-G mixing amplitude (at $\Delta E_{HG}=|E^0_G-E^0_H|\simeq 0.5$GeV):
$$
\begin{array}{c}
\lambda_{HG}=|\frac{V_{HG}}{E^0_M-E^0_G}|\simeq 0.18 \simeq 18\%,\\
\end{array}
$$
and to very small amplitude $\lambda_{MG}$ of MG mixing:
$$
\lambda _{MG}=\lambda_{MH}\lambda_{HG}\simeq0.045\simeq4.5\%
$$ 

$\bullet$ The value of $\Delta E^0_n=\Delta m_n=\sum\limits_{k\ne n}\frac{|V_{nk}|^2}{E_n^0-E_k^0}$ in Eq. (7) gives in CM system of any particle the correction to its mass $m_n^0$ due to the mixings with  higher Fock states. These corrections to meson, to hybrid and to glueball masses prove to be very small:
 $$
\begin{array}{c}
\Delta m_M\simeq -\Delta m_H=\frac{|V_{MH}|^2}{E^0_M-E^0_H}\simeq -\frac{(0.20eV)^2}{0.8GeV}\simeq -50MeV,\\ \\
\Delta m_G=\frac{|V_{HG}|^2}{E^0_G-E^0_H}\simeq\frac{(0.091GeV)^2}{\pm 0.5GeV}\simeq\pm 16.5MeV
\end{array}
$$
where the energy gups $\Delta E ^o_{MH}$ and $\Delta E ^o_{HM}$ were put for estimations to be equal to some fraction of $1GeV$,$(\Delta E^o_{GH} \simeq 0.5 GeV,\Delta E_{MH}\simeq0.8$ GeV)

There exists a danger that highly exiting meson states with $E^0_M\sim 1$GeV can enhance significantly these estimation for $\Delta m_H$.\\

{\bf 5. Summary.}\\

$\bullet$ It is argued that gluon field can be represented as the sum of continuous part $B_\mu(x)$ leading to gluon string creation and to confinement of valent (constituent) quarks and gluons inside hadrons in a small volume $V\sim R^3_0,\; R_0\sim0.5-0.6$fm and the part $a_\mu(x)$ which produce and absorb these valent quarks and gluons. Due to the confinement these valent colour particles acquire not small mass $\sim 1/R_0\sim 0.3-0.4$GeV. Also gluon string itself has about the same mass.

$\bullet$ Therefore each extra gluon, or pair of  quarks $\bar qq$ bound by gluon strings lead to enhancement of hadron mass to about of $|\Delta E| \simeq (0.8-1.0)$GeV

$\bullet$ In the quantum mechanical (QM) perturbation theory (PT) this large energy gap leads to rapidly convergent series as the PT matrix elements of QCD interaction (2) $v'_{QCD}(a_\mu)=g_s\int(\bar q \gamma_\mu t_\alpha q)a_\mu^\alpha(x)d^3x$ connecting these sectors (like meson $\bar qq$, hybrid $\bar qqg$, glueball $gg$ with different number of valent gluons and $\bar q q$ paires) are small: $V_{MH}\simeq0.2$GeV, $V_{HG}\simeq0.1$GeV, $V_{MG}\simeq 0.02$GeV. 

These figures and those given above were obtained in the first approximation in expansion of integrand in Eq.(18) in powes of $\langle (1-t/b)^n\rangle$. It converges only asymtotically (up to $n\leq 3-4$) and the odd powers of $n$ gives to $J(a,b)$ the negative contribution. The next two terms in Eq.(18) up to $n\leq 3$ enhance slightly the value of $MH$ and $HG$ mixings changing nothing in princiole (as $V_{MH}$ changes from $\sim 0.20$GeV only up to $0.26$GeV while $V_{HG}$ encreases from $0.09$GeV for $n\leq 1$ up to $0.013$GeV at $n\leq 3$ in Eqs.(18)--(20).  

$\bullet$ As a result the QCD problem of colourless bound states can have a simple exact solution in the form of convergent usual QM perturbation series which converges rapidly just due to the confinement effect.


\vspace*{0cm}
$\bullet$ The developed logic can have a few dangerous underwater stones: in the suggested "stript" relativistic approach the small dependence of functions $\Phi_M(r),\:\Phi_G(r),\:\Phi_H(\vec r,\vec r_3)$ on quark spin variables was not taken accurately into account. The more accurate approach is now in progress.\\[1cm]


\begin{center}
{\bf Acknowledgement}
\end{center}
A larger part of this paper is based on results of Yu.Simonov and his group. The author thanks him and his colleagues and in particular, Yulia Kalashnikova and B.Kerbikov for help and useful discussions.\\[1cm]

\begin{center}
{\bf References.}
\end{center}
[1] Yu.A.Simonov Lectures on QCD at XVII school of Physics, Lisbon, 1999.

\noindent [2] V.L.Morgunov, V.I.Shevchenko, Yu.A.Simonov Phys. of Atom. Nuclei {\bf 61}(1998)219.

\noindent [3] V.D.Mur, V.S.Popov, Yu.A.Simonov and V.P.Yurov JETP {\bf 105}(1998)3--27.

\noindent [4] R.Calogero, and Yu.A.Simonov Phys.Rev. {\bf 169}(1968)789. 

\noindent [5] A.LeYaouane, L.Oliver, O.P\^ene, J.-C.Raynal, S.Ono Zs.Phys C, Particles and Fields 
\hspace*{0.6cm}{\bf C28}(1985)309, also Phys.Rev. {\bf D 29}þ

\noindent [6] E.Gubankova, C.--R.Ji, S.Cotanch, hep--ph/0003289, 30/III 2000.

\noindent [7] E.Gubankova, C.--R.Ji, S.Cotanch, hep--ph/9908331 v.2 23/III 1999.


\begin{center}
\epsfig{figure=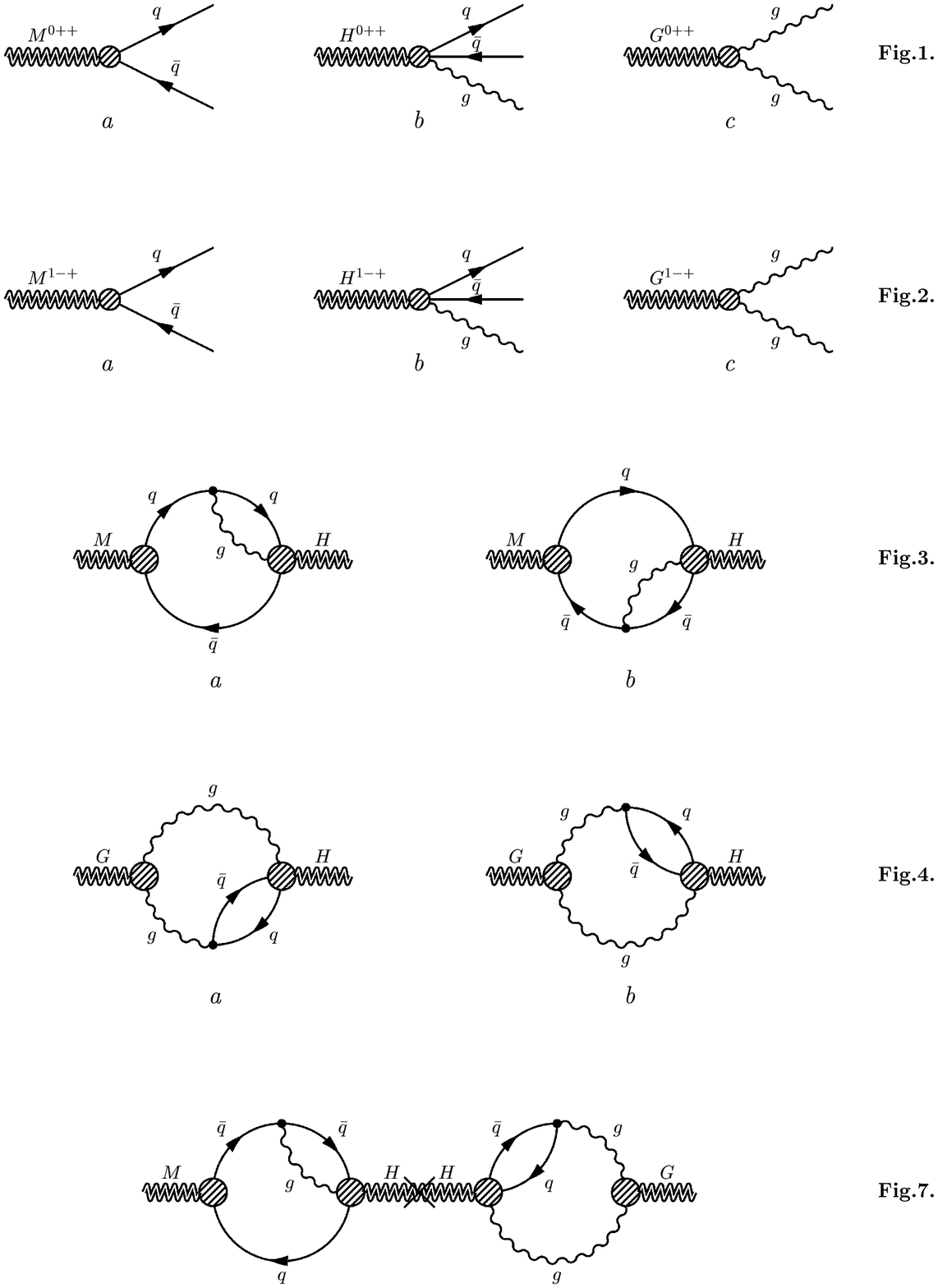, height=230mm, width=170mm, clip=}
\end{center}

\newpage

\vspace*{2cm}
\begin{center}
\epsfig{figure=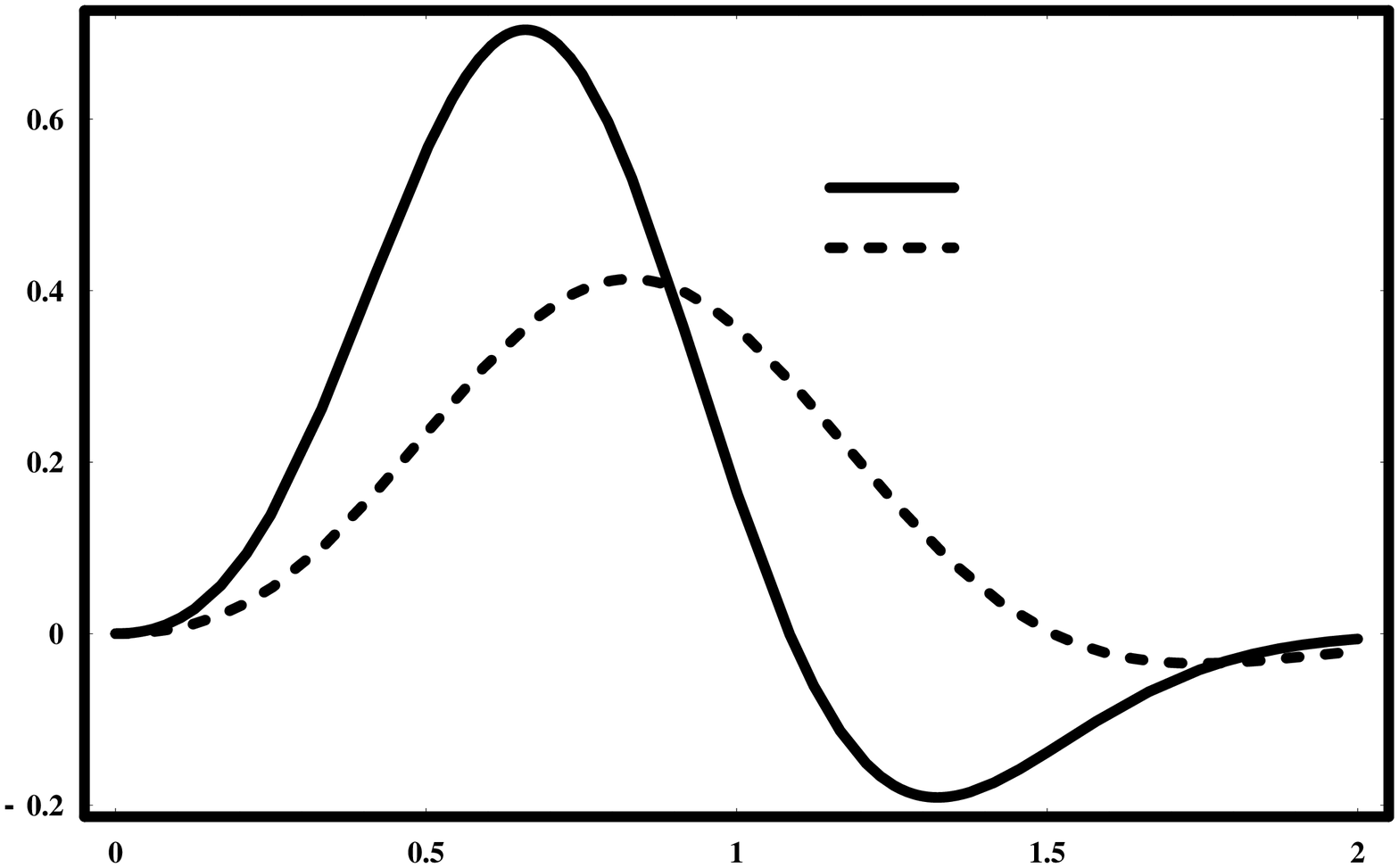, height=110mm, width=170mm, clip=}
\put(-49.5,87){\large\bf{$a=2.791,\;b=0.775$}}
\put(-49.5,78){\large\bf{$\tilde a=1.444,\; \tilde b=1.078$}}
\put(-90,90){\large\bf{$a)$}}
\put(-50,40){\large\bf{$b)$}}
\vspace*{1cm}
\large\bfseries{Fig.6a,b}
\end{center}

\end{document}